\documentclass[twocolumn,aps,prl,superscriptaddress,nofootinbib,showpacs]{revtex4-1}
\usepackage{graphicx}
\usepackage{color}
\graphicspath{{figures/}{fig/}}
\usepackage{amsmath}
\usepackage{bm}
\usepackage{slashed}
\usepackage{epsfig}
\usepackage{amsfonts}
\usepackage{epstopdf}
\usepackage{hyperref}
\usepackage{bbm}
\usepackage{textcomp}
\newtheorem{alg}{Algorithm}
\newcommand{\sect}[1]{{\it \textbf{#1} --- }}
\newcommand{\ui}{\mathrm{i}}
\newcommand{\ud}{\mathrm{d}}
\begin{document}
\title{Determining arbitrary Feynman integrals by vacuum integrals}

\author{Xiao Liu}
\email{xiao6@pku.edu.cn}
\affiliation{School of Physics and State Key Laboratory of Nuclear Physics and
Technology, Peking University, Beijing 100871, China}
\author{Yan-Qing Ma}
\email{yqma@pku.edu.cn}
\affiliation{School of Physics and State Key Laboratory of Nuclear Physics and
Technology, Peking University, Beijing 100871, China}
\affiliation{Center for High Energy Physics, Peking University, Beijing 100871, China}
\affiliation{Collaborative Innovation Center of Quantum Matter,
Beijing 100871, China}

\date{\today}

\begin{abstract}
By introducing an auxiliary parameter, we find a new representation for Feynman integrals, which defines a Feynman integral by analytical continuation of a series containing only vacuum integrals. The new representation therefore conceptually translates the problem of computing Feynman integrals to the problem of performing analytical continuations. As an application of the new representation, we use it to construct a novel reduction method for multiloop Feynman integrals, which is expected to be more efficient than the known integration-by-parts reduction method. Using the new method, we successfully reduced all complicated two-loop integrals in the $gg\to HH$ process and $gg\to ggg$ process.
\end{abstract}

\maketitle
\allowdisplaybreaks

%%%%%%%%%%%%%%%%%%%%%%%%%%%%%%%%%%%%%%%%%%%%%%%%%
\sect{Introduction}
Computation of Feynman loop integrals is in the heart of modern physics, which is important both for testing the particle physics standard model and for discovering new physics. A good method to compute one-loop integrals was proposed as early as the 1970s, the strategy of which is to first express scattering amplitudes in terms of linear combinations of master integrals (MIs) and then compute these MIs \cite{Passarino:1978jh,tHooft:1978jhc,vanOldenborgh:1989wn}. Based on this method, one can compute one-loop scattering amplitudes systematically and efficiently if the number of external legs is no more than 4. With further improvement of the traditional tensor reduction \cite{Denner:2005nn} and the development of unitarity-based reduction  \cite{Britto:2004nc,Ossola:2006us,Giele:2008ve}, computation of multileg one-loop scattering amplitudes is also a solved problem right now.

Yet, about 40 years later, it is still a challenge to compute multiloop integrals, even for two-loop integrals with four external legs. The mainstream approach to calculate multiloop integrals in the literature is similar to that at one-loop level, by first reducing Feynman integrals to MIs and then calculating these MIs. However, both of the two steps are much harder to achieve than the one-loop case.

Although compact and explicit expressions for one-loop MIs can be easily obtained \cite{tHooft:1978jhc,vanOldenborgh:1989wn},
the computation of multiloop MIs is very challenging. There are many methods in the literature to compute multiloop MIs, such as the sector decomposition~\cite{Binoth:2000ps,Heinrich:2008si,Borowka:2017idc,Carter:2010hi,Borowka:2012yc,Borowka:2015mxa,Bogner:2007cr,Smirnov:2008py,Smirnov:2009pb,Smirnov:2013eza,Smirnov:2015mct,Gluza:2010rn}, Mellin-Barnes representation ~\cite{Usyukina:1975yg,Smirnov:1999gc,Tausk:1999vh,Heinrich:2004iq,Czakon:2004wm,Czakon:2005rk,Anastasiou:2005cb,Gluza:2007rt,Gluza:2010rn,Dubovyk:2015yba,Dubovyk:2016ocz,Dubovyk:2016aqv,Gluza:2016fwh,Dubovyk:2017cqw,Sidorov:2017aea,Prausa:2017frh}, and the differential equation method~\cite{Kotikov:1990kg,Bern:1992em,Remiddi:1997ny,Gehrmann:1999as,Henn:2013pwa,Lee:2014ioa}, but none of them provides a fully satisfactory solution.
In Ref.~\cite{Liu:2017jxz}, we proposed a systematic and efficient method to calculate multiloop MIs by constructing and numerically solving a system of ordinary differential equations (ODEs). The differential variable, say $\eta$, is an auxiliary parameter introduced to all Feynman propagators. With the ODEs, physical results at $\eta=0^+$ are fully determined by boundary conditions chosen at $\eta=\infty$, which can be obtained almost trivially. Therefore, MIs can be treated as special functions in our method, but it relies on a good reduction method to set up ODEs.

Reduction of multiloop integrals is an even harder problem. Significantly different from the one-loop case, propagators in a multiloop integral are usually not enough to form a complete set to expand all independent scalar products, either between a loop momentum and an external momentum or between two loop momenta. As a consequence, the unitarity-based multiloop reduction \cite{Gluza:2010ws,Kosower:2011ty,Mastrolia:2012wf,Badger:2012dp,Mastrolia:2013kca,Larsen:2015ped,Ita:2015tya,Badger:2016ozq,Mastrolia:2016dhn,Abreu:2017idw,Abreu:2017xsl} has difficulty fully reducing scattering amplitudes. Although the integration-by-parts (IBP) reduction~\cite{Chetyrkin:1981qh,Laporta:2001dd,Studerus:2009ye,Lee:2012cn,Smirnov:2014hma} is general enough to reduce any scattering amplitude to MIs, the incompleteness of multiloop propagators makes it hard to generate efficient reduction relations. Currently, IBP reduction is mainly based on Laporta's algorithm \cite{Laporta:2001dd}, which usually generates huge number of linear equations which is extremely hard to solve for multiscale problems. For example, it cannot give a complete reduction for Higgs pair hadroproduction at two-loop order~\cite{Borowka:2016ehy}. Improvements for IBP reduction method can be found in \cite{vonManteuffel:2014ixa,Boehm:2017wjc} and references therein.

Inspired by our previous work~\cite{Liu:2017jxz}, in this article we construct a novel method to compute Feynman loop integrals. The key observation is that, after introducing the auxiliary parameter $\eta$, any Feynman integral can be defined as the analytical continuation of a calculable asymptotic series, which contains only simple vacuum integrals. The series can thus be thought of as a new representation of the Feynman integral.
Based on the new representation,  we construct an efficient reduction method for multiloop integrals. We demonstrate the correctness and efficiency of our reduction method with two cutting-edge examples.

%%%%%%%%%%%%%%%%%%%%%%%%%%%%%%%%%%%%%%%%%%%%%%%%%
\sect{A new representation for Feynman integrals}
Following Ref.~\cite{Liu:2017jxz}, we introduce a dimensionally regularized $L$-loop Feynman integral with an auxiliary parameter $\eta$,
\begin{align}\label{eq:int}
{\cal M}(D,\vec{s}\,,\eta)\equiv\int\prod_{i=1}^{L}\frac{\ud^D\ell_i}{\ui\pi^{D/2}}\prod_{\alpha=1}^{N}\frac{1}{(\mathcal{D}_\alpha+\ui\eta)^{\nu_{\alpha}}}\, ,
\end{align}
where $D$ is the spacetime dimension, $\mathcal{D}_\alpha\equiv q_\alpha^2-m_\alpha^2$ are usual inverse Feynman propagators with $m_\alpha$ being corresponding masses and $q_\alpha$ being linear combinations of loop momenta $\ell_i$ and external momenta $p_i$, $\vec{s}=(s_1,\ldots,s_r)$ are independent kinematic variables (including mass parameters) in the problem, and $\nu_\alpha$ are powers of propagators whose dependence is suppressed in the left-hand side of the equation. The physical integral that we want to get is
\begin{align}
{\cal M}(D,\vec{s}\,,0^+)\equiv \lim_{\eta\to0^+} {\cal M}(D,\vec{s}\,,\eta),
\end{align}
with $0^+$ defining the causality of Feynman amplitudes. %In the following, we will suppress the dependence on $\{\nu_\alpha\}$ whenever it does not introduce any confusion.

The study in Ref.~\cite{Liu:2017jxz} shows that, as $\eta\to\infty$, there is only one integration region for  ${\cal M}(D,\vec{s}\,,\eta)$, where all components of loop momenta are of the order of $|\eta|^{1/2}$. Therefore, all propagators can be expanded like
\begin{align}\label{eq:expandPro}
&\frac{1}{[(\ell +p)^2-m^2+\ui \eta]^\nu}\nonumber\\
=& \frac{1}{(\ell^2+\ui \eta)^\nu}\sum_{j=0}^\infty\frac{(\nu)_j}{j!} \left(-\frac{2 \ell\cdot p+p^2-m^2}{\ell^2+\ui \eta}\right)^j\,,
\end{align}
where $\ell$ is a linear combination of loop momenta $\ell_i$, $p$ is a linear combination of external momenta $p_i$, and $(\nu)_j\equiv\nu(\nu+1)\cdots(\nu+j-1)$ is the Pochhammer symbol. After the expansion, all external momenta and masses are not present in denominators anymore, thus each term of the expansion can be interpreted as vacuum integrals with equal internal squared masses $-\ui \eta$.
Inserting Eq.~\eqref{eq:expandPro} into Eq.~\eqref{eq:int} and rescaling all loop momenta by $\eta^{1/2}$, we can obtain an asymptotic expansion around $\eta=\infty$,
\begin{align}\label{eq:expandM}
{\cal M}&(D,\vec{s}\,,\eta)=\eta^{LD/2-\sum_\alpha \nu_\alpha}\sum_{\mu_0=0}^\infty\eta^{-\mu_0}\mathcal{M}^\mathrm{bub}_{\mu_0}(D,\vec{s}\,)\,,
\end{align}
where the superscript ``bub" means vacuum bubble integrals.
Although asymptotic expansion of Feynman integrals itself is not new~\cite{Smirnov:1994tg,Smirnov:2002pj}, the novelty here is that our expansion is with respect to an auxiliary parameter introduced by hand and thus can be applied to any problem. In Eq.\eqref{eq:expandM}, $\mathcal{M}^\mathrm{bub}_{\mu_0}(D,\vec{s}\,)$ consist of vacuum integrals  with equal internal squared masses $-\ui$, which can be easily reduced to vacuum MIs denoted by $\{I_{L,1}^{\mathrm{bub}}(D),\ldots,I_{L,B_L}^{\mathrm{bub}}(D)\}$. Here $B_L$ is the total number of  $L$-loop equal-mass vacuum MIs, with $B_1=1$, $B_2=2$, $B_3=5$ and so on. Thus, after the vacuum reduction, we have the decomposition
\begin{align}\label{eq:expandMbub}
\mathcal{M}^\mathrm{bub}_{\mu_0}(D,\vec{s}\,)=\sum_{k=1}^{B_L}I^{\mathrm{bub}}_{L,k}(D)\sum_{\vec{\mu}\in \Omega^r_{\mu_0}}C_k^{\mu_0\ldots\mu_r}(D) s_1^{\mu_1}\cdots s_r^{\mu_r}\,,
\end{align}
where $\vec{\mu}$ is a $r$-dimensional vector in $\Omega^r_{\mu_0}\equiv\{\vec{\mu}\in\mathbb{N}^r|\,\mu_1+\cdots+\mu_r=\mu_0\}$, and $C_k^{\mu_0\ldots\mu_r}(D)$ are rational functions of $D$.
%The expansion \eqref{eq:expandM} can be alternatively obtained from Feynman parametric representation.

As vacuum MIs can be easily calculated~\cite{Davydychev:1992mt,Broadhurst:1998rz,Kniehl:2017ikj,Schroder:2005va,Luthe:2015ngq,Luthe:2017ttc}, the series \eqref{eq:expandM} defines an analytical function around $\eta=\infty$, which therefore determines ${\cal M}(D,\vec{s}\,,\eta)$ for any value of $\eta$ based on analytical continuation. Especially,  the desired physical value at $\eta=0^+$ is fully determined. As a result, the expression  \eqref{eq:expandM} can be thought as a new series representation of ${\cal M}(D,\vec{s}\,,0^+)$. Then the problem of computing Feynman integrals is translated to the problem of performing analytical continuations. This conceptual change of interpretation of Feynman integrals may both deepen our understanding of scattering amplitudes and result in powerful methods to compute scattering amplitudes.

Beginning in the next section, we are devoted to constructing a powerful reduction method for Feynman integrals based on the new representation. The reduction method can not only reduce any Feynman integral to MIs $\vec{I}(D,\vec{s}\,,\eta)$ (note that there are more MIs after the introduction of $\eta$), but also set up a system of ODEs for these MIs
\begin{align}\label{eq:DE}
\frac{\partial}{\partial \eta} \vec{I}(D,\vec{s}\,,\eta)= A(D,\vec{s}\,,\eta)\vec{I}(D,\vec{s}\,,\eta).
\end{align}
Then the analytical continuation from $\eta=\infty$ to $\eta=0^+$ can be realized by solving the ODEs \cite{Liu:2017jxz}.

%%%%%%%%%%%%%%%%%%%%%%%%%%%%%%%%%%%%%%%%%%%%%%%%%
\sect{Reduction relations from the new representation}
An important property of Feynman loop integrals is that the number of MIs is {\it finite}~\cite{Smirnov:2010hn}. More precisely, for loop integrals constructed from any given set of propagators, there exists a {\it finite} set of loop integrals called MIs
%(called MIs, which can be found easily~\cite{Georgoudis:2016wff})
so that all other loop integrals can be expressed as linear combinations of them, with coefficients being rational functions of kinematic variables and spacetime dimension.
The reduction is to find relations among loop integrals and eventually express all loop integrals by MIs.
%In other words, loop integrals with given set of propagators form a finite-dimensional linear space.
%Traditionally, people use the IBP reduction based on Laporta's algorithm~\cite{Laporta:2001dd} to reduce their desired integrals to master integrals. This approach, however, might result in large coupled linear systems  when the problems are too complicated. And the efficiency of the Gaussian elimination will extremely decrease as the number of scales grows. So, traditional IBP reduction is not good at solving multi-scale problems.
%Our method, on the other hand, can handle these ``traditionally difficult'' problems. We'll illustrate the main idea in next few paragraphs.
%In the following discussions, we will suppress the dependence on $D$, $\vec{s}$ and $\eta$ in loop integrals whenever it does not introduce any confusion.

Let us first study how to find relations among a given set of loop integrals using the new representation.
Suppose we have a set of integrals $G=\{\mathcal{M}_1,\ldots,\mathcal{M}_{n}\}$. Linear relations among them can be written as
\begin{align}\label{eq:relation}
\sum_{i=1}^{n} {Q_i(D,\vec{s}\,,\eta)}\, \mathcal{M}_i(D,\vec{s}\,,\eta)=0\,,
\end{align}
where $Q_i$ are homogeneous polynomials of $\eta$ and kinematic variables $\vec{s}$. We denote the mass dimension of $\mathcal{M}_i$ by $\mathrm{Dim}(\mathcal{M}_i)$ and the degree of $Q_i$ by $d_i$, which are constrained by
\begin{align}\label{eq:dimension}
2d_1+\mathrm{Dim}(\mathcal{M}_1)=\cdots=2d_{n}+\mathrm{Dim}(\mathcal{M}_{n})\,.
\end{align}
Therefore, there is only one degree of freedom in $\{d_i\}$, which can be chosen as $d_{\text{max}}=\text{max}\{d_i\}$.

For any given $d_{\text{max}}\geq 0$, we can expand each $Q_i(D,\vec{s}\,,\eta)$ as
\begin{align}\label{eq:expandQ}
Q_i(D,\vec{s}\,,\eta) = \sum_{(\lambda_0,\vec{\lambda})\in\Omega^{r+1}_{d_i}} Q_i^{ \lambda_0 \ldots \lambda_r}(D) \, \eta^{\lambda_0} s_1^{\lambda_1}\cdots s_r^{\lambda_r},
\end{align}
where $Q_i^{ \lambda_0 \ldots \lambda_r}(D)$ are rational functions of $D$ to be determined [note that by definition $Q_i(D,\vec{s}\,,\eta)\equiv0$ if $d_i<0$]. As the series \eqref{eq:expandM} fully determines all analytical functions ${\cal M}_i$, it certainly also determines the relations among them in Eq.~\eqref{eq:relation}.
To determine the unknown coefficients $Q_i^{ \lambda_0 \cdots \lambda_r}(D)$, we substitute Eqs.~\eqref{eq:expandM},\eqref{eq:expandMbub}, and \eqref{eq:expandQ} into Eq.~\eqref{eq:relation} and then expand it in terms of monomials of $I^{\mathrm{bub}}_{L,k}(D)$, $\eta$, and $\vec{s}$, which gives
\begin{align}\label{eq:matching}
\sum_{k,\rho_0,\vec{\rho}} f_k^{\rho_0\ldots\rho_r}\,I^{\mathrm{bub}}_{L,k}(D)~\eta^{\rho_0}s_1^{\rho_1}\cdots s_r^{\rho_r}=0\,,
\end{align}
where $f_k^{\rho_0\ldots\rho_r}$ are linear functions of $Q_i^{ \lambda_0 \cdots \lambda_r}(D)$. As  $I^{\mathrm{bub}}_{L,k}(D)~\eta^{\rho_0}s_1^{\rho_1}\cdots s_r^{\rho_r}$ are independent of each other, their coefficients must vanish, which results in a system of linear equations
\begin{align}\label{eq:LEQs}
f_k^{\rho_0\ldots\rho_r}=0,~~~\text{for each $k$, $\rho_0$, $\ldots$, $\rho_r$. }
\end{align}
By calculating the series \eqref{eq:expandM} to sufficiently high order in $1/\eta$, we can generate enough linear equations to constrain the solution space of $Q_i^{ \lambda_0 \cdots \lambda_r}(D)$. In practice, we find that it is sufficient if the number of linear equations is larger than the number of unknown coefficients $Q_i^{ \lambda_0 \cdots \lambda_r}(D)$ by $30\%$. Once the solution space is obtained, it provides us with all relations among the integral set $G$ with given $d_{\text{max}}$.
\footnote{In practice, we solve the solution space with some chosen
	non-special values of $D$, e.g., $D=1867/5281$, and then reconstruct the $D$-dependence for the solutions. Finite field technique~\cite{vonManteuffel:2014ixa,Peraro:2016wsq} is also used to speed up the calculation.
%We first reconstruct their dependence on $D$ in some prime fields, and then we preform the rational reconstruction to determine the rational coefficients.
}

Now let us consider two sets of integrals, $G_1$ and $G_2$, with the condition that integrals in $G_2$ are all simpler than integrals in $G_1$. Assuming that $G_1$ can be reduced to $G_2$, we provide an algorithm to find out relations to realize this reduction.
\begin{alg}	\label{al:1}
	\begin{enumerate}
		\item Let $G=\{G_1,G_2\}$ and $d_{\text{max}}=0$.
		\item Generate and solve the linear equations in \eqref{eq:LEQs} to obtain all possible relations. \label{al:calc}
		\item If the obtained relations are enough to express $G_1$ in terms of $G_2$, stop; otherwise, increase $d_{\text{max}}$ by 1 and go to step \ref{al:calc}.
	\end{enumerate}
\end{alg}
According to our assumption, the iteration must terminate after finitely many steps, because $Q_i(D,\vec{s}\,, \eta)$ are polynomials with finite degree.
%Therefore, this algorithm is executable and can be realized automatically on computer systems.
Our algorithm is constructed to {\it search} for as simple as possible relations.

We emphasize that, although they are determined by the region  $\eta\to\infty$, the obtained relations are correct for any value of $\eta$ and thus can be used to reduce physical Feynman loop integrals.

%%%%%%%%%%%%%%%%%%%%%%%%%%%%%%%%%%%%%%%%%%%%%%%%%
\sect{Reduction scheme}
To reduce a given integral to simpler integrals, we still need to choose $G_1$, which includes the given integral as an element, and $G_2$, which includes only simpler integrals and can express all integrals in $G_1$. There are many possible choices, but a good choice should satisfy the following: (1) Relations among $\{G_1,G_2\}$ are simple, so that they can be easily found using algorithm \ref{al:1}; (2) the number of integrals in $G_1$ is not too large, so that one can efficiently solve the obtained relations to reduce $G_1$.
To simplify our discussion, in the following we only consider the reduction of scalar integrals,\footnote{Note that reduction of scalar integrals is already general enough. First, any tensor integral can be easily expressed as scalar integrals in higher spacetime dimension ~\cite{Davydychev:1991va,Tarasov:1996br,Tarasov:1997kx}. Second, recurrence relations to relate higher spacetime-dimension integrals to lower spacetime-dimension integrals~\cite{Tarasov:1996br,Lee:2009dh} can be obtained by reduction relations of scalar integrals.} which means integrals with no numerator in the integrand.

Let us begin with introducing some notations. For a given set of propagators, a scalar integral can be denoted by its powers of corresponding propagators $\vec{\nu}=(\nu_1,\ldots,\nu_N)$ $(\nu_i\geq0)$.
A {\it sector} is a set of integrals that have exactly the same 0s in the powers, e.g., $(5,1,0,3)$ and $(7,8,0,9)$ are in the same sector.
We introduce two sets of operators $\mathbf{m^+}$ and $\mathbf{m^-}$, with positive integer $m$. If $m>1$, we define $\mathbf{m^{\pm}}=\mathbf{(m-1)^{\pm}}\mathbf{1^{\pm}}$. $\mathbf{1^+}$ ($\mathbf{1^-}$) is defined so that, when applying it on an integral $\vec{\nu}$, it generates all integrals with one nonvanishing $\nu_i$ increased (decreased) by 1. For example, $\mathbf{1^+}(5,1,0,3)=\{(6,1,0,3),(5,2,0,3),(5,1,0,4)\}$ generating integrals in the same sector, and $\mathbf{1^-}(5,1,0,3)=\{(4,1,0,3),(5,0,0,3),(5,1,0,2)\}$ generating integrals either in the same sector or in subsectors. Note that $\mathbf{m^+}\mathbf{n^-}\neq \mathbf{n^-}\mathbf{m^+}$, which can be easily verified.

To figure out a good choice of $\{G_1,G_2\}$ at multiloop level, we should first take a look at the one-loop case to see what we can learn. In this case, it is well known that quite simple relations can be obtained to reduce integrals $G_1=\mathbf{1^+}\vec{\nu}$ to simpler integrals $G_2=\mathbf{1^-1^+}\vec{\nu}$~\cite{Duplancic:2003tv}. This reduction is possible because there is only one MI in each sector at one-loop level, and thus even integrals like $\mathbf{1^+}(1,1,0,1)$ are fully reducible.

As a natural generalization of one-loop strategy, we propose to reduce integrals $G_1=\mathbf{m^+}\vec{\nu}$ at multiloop level, where $m$ is usually larger than 1 because there is usually more than one MI in each sector. The smallest allowed value of $m$, which guarantees that all integrals $\mathbf{m^+}\vec{\nu}$ are reducible to simpler integrals, can be found out within our method. Alternatively, it can be easily determined by investigating the distribution of MIs in the sector containing $\vec{\nu}$, because finding out MIs is a simple problem~\cite{Georgoudis:2016wff}. In the following examples, we find $m=2$ or $3$, thus there are only dozens of integrals in $G_1$.

A possible generalization for the set of simpler integrals is then $G_2=\{\mathbf{1^-m^+},\mathbf{2^-m^+},\ldots,\mathbf{m^-m^+}\}\vec{\nu}$, or its subset $G_2=\{\mathbf{1^-m^+},\mathbf{1^-(m-1)^+},\ldots,\mathbf{1^-1^+}\}\vec{\nu}$. In the following examples, we use the latter choice and find that it can already result in not too complicated relations.

There are exceptions where a reducible integral cannot be included in any fully reducible set $\mathbf{m^+}\vec{\nu}$. This is harmless because it only happens when this integral and some MIs have the same $|\vec{\nu}|$. We can either put it in a partially reducible set, or simply treat it as a redundant MI.

With the above strategy, we can express any reducible integral as linear combinations of simpler integrals. Then by iteration, we can reduce any integral to MIs.

Therefore, we realize a step-by-step reduction scheme for multiloop integrals, which is similar to the one-loop case. Comparing with the traditional IBP reduction method, an advantage of our method is that we never encounter large coupled linear systems. As a result, the computation complexity of numerically solving the obtained reduction relations to reduce $N$ integrals to MIs is $O(N)$, rather than $O(N^3)$ in the fully coupled case.%, no matter how many kinematic variables are involved in the process.

%Let us denote $\vec{I}(D,\vec{s},\,\eta)$ as the vector of a complete set of $n$ MIs. As $\frac{\partial}{\partial\eta}\vec{I}(D,\vec{s},\,\eta)$ are special loop integrals, the reduction method described above can also express them in terms of linear combinations of $\vec{I}(D,\vec{s},\,\eta)$. Therefore, we obtain a system of ODEs,
%\begin{align}\label{eq:DE}
%\frac{\partial}{\partial\eta}\vec{I}(D,\vec{s},\,\eta)=A(D,\vec{s},\,\eta)\vec{I}(D,\vec{s},\,\eta)\, ,
%\end{align}
%where  $A(D,\vec{s},\,\eta)$ is the calculable $n\times n$ coefficient matrix. With the ODEs, analytical continuation of MIs from $\eta=\infty$ to $\eta=0^+$ can be obtained straightforwardly by numerically solving the ODEs with BCs chosen at $\eta=\infty$. The process is well-studied mathematically, and final results can be obtained efficiently to high precision~\cite{Liu:2017jxz}.

%We eventually find that all MIs, and thus arbitrary loop integrals, can be determined unambiguously  by the series representation, which basically  involves only vacuum integrals.

\sect{Examples}
To test the power of our new reduction method, we apply it to two cutting-edge processes.
The first example is a two-loop four-scale on-shell scattering $gg\rightarrow HH$ with a top quark loop. The second one is two-loop five-gluon on-shell scattering process with five independent scales. Integrals of the two examples have not been fully reduced by the traditional IBP reduction method~\cite{Borowka:2016ehy}.

We have tried integrals in many sectors, and found all of them can be easily reduced using our method. Three conclusions based on our test are as follows. First, as expected the more external legs the harder the reduction is.  Second, reduction for nonplanar integrals is typically harder than planar integrals, which may be caused by the fact that there are usually more MIs in nonplanar sectors. Finally, suppose that $\vec\nu$ and $\vec e$ belong to the same sector and that $\vec e$ is the simplest integral, which includes only single power propagators, then the reduction for $\mathbf{m^+}\vec{e}$ is usually more difficult than the first-step reduction for $\mathbf{m^+}\vec{\nu}$. This can be understood because the set $\{\mathbf{1^-m^+},\ldots,\mathbf{1^-1^+}\}\vec{e}$ contains fewer integrals in the leading sector and thus has fewer flexible relations to express desired reducible integrals. With these observations, we then mainly discuss potentially difficult integrals.

\begin{figure}[htb]
\begin{center}
\includegraphics[width=0.9\linewidth]{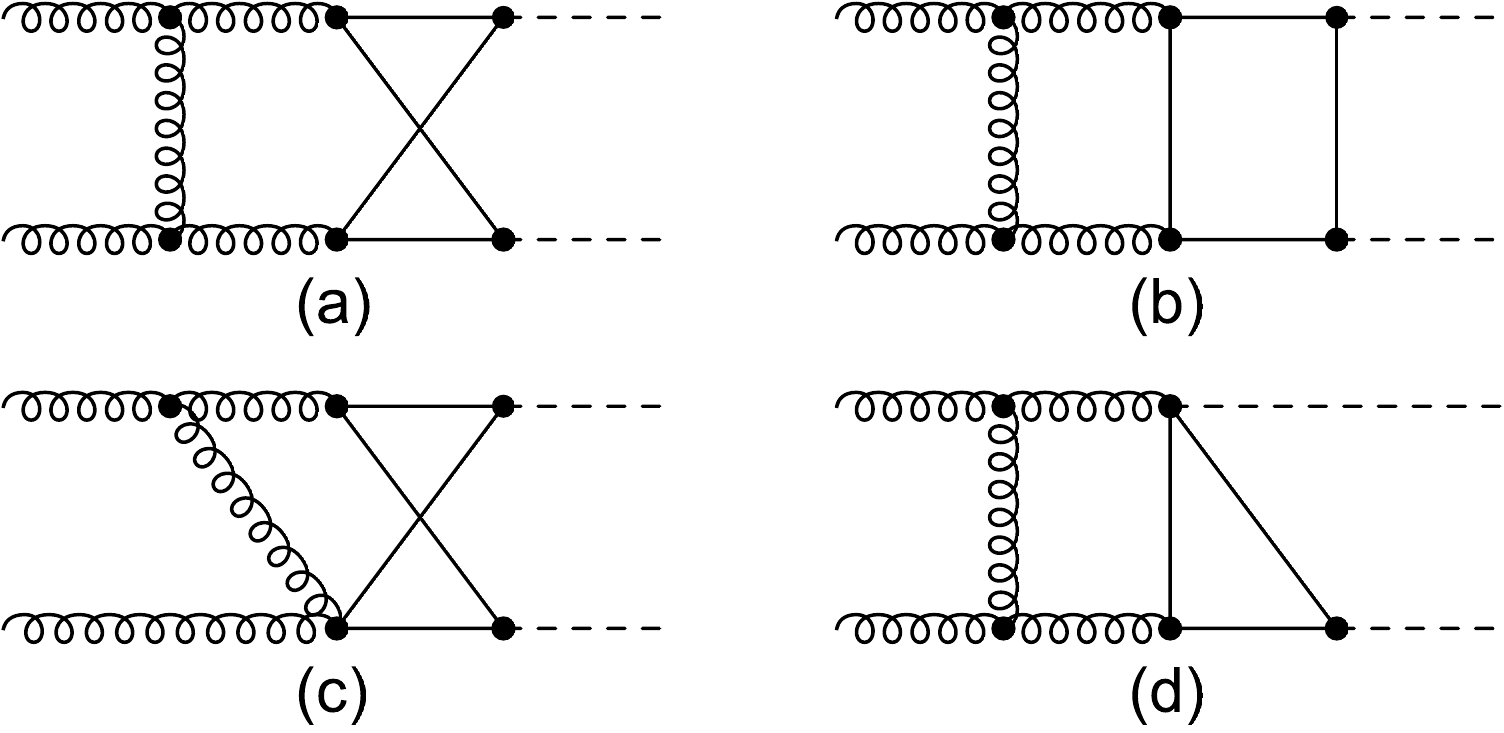}
\caption{\label{fig:ggToHH}
Some typical diagrams in the $gg\rightarrow HH$ process. Diagrams (c) and (d) are obtained from (a) and (b) by shrinking a gluon and a top quark line, respectively.}
\end{center}
\end{figure}

\begin{figure}[htb]
	\begin{center}
		\includegraphics[width=0.9\linewidth]{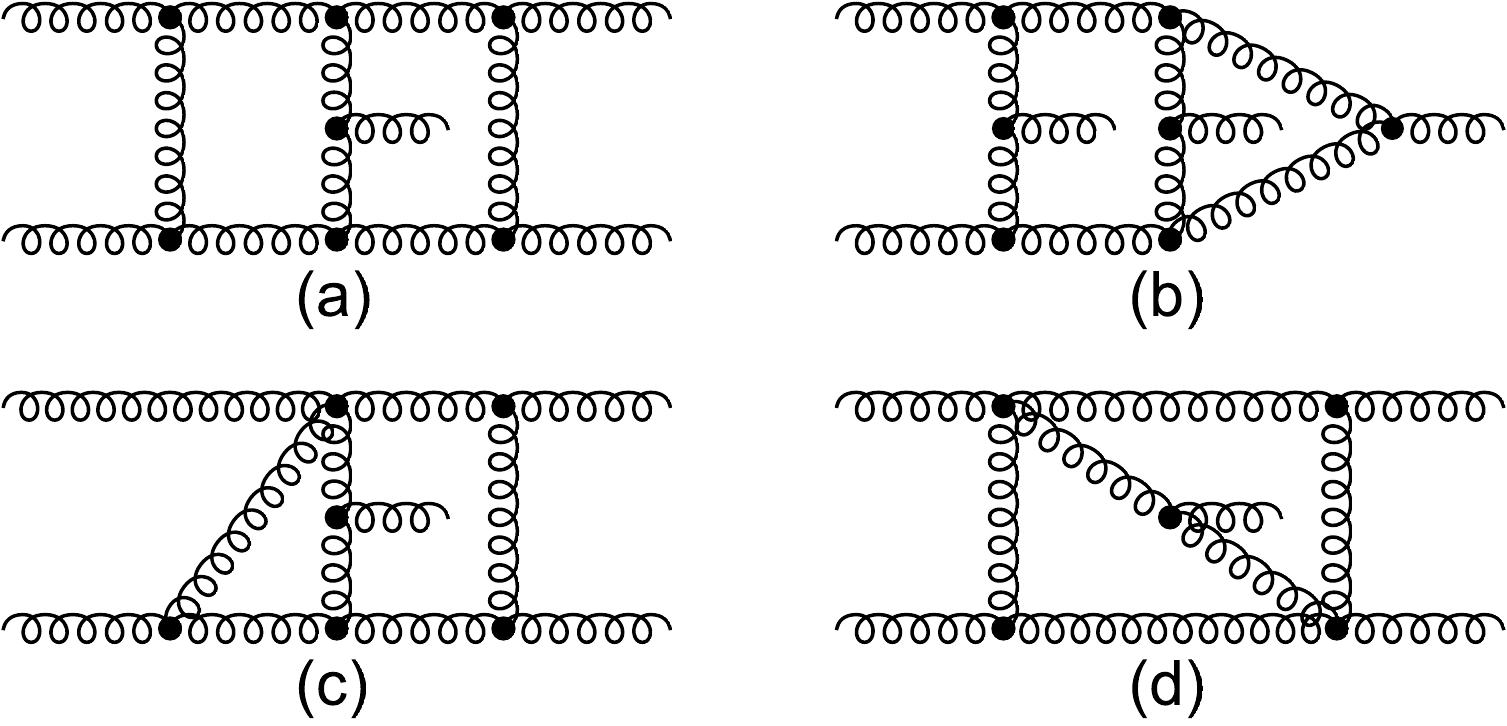}
		\caption{\label{fig:5g}
			All nonplanar five-leg diagrams for the five-gluon process.}
	\end{center}
\end{figure}

For the first example, some typical diagrams are shown in Fig.\ref{fig:ggToHH}, and corresponding reduction information is summarized in Table \ref{tab:time}. Here, in addition to two potentially difficult nonplanar diagrams, we also show two planar diagrams to compare with. We find that the six-propagator nonplanar sector Fig.\ref{fig:ggToHH}(c) is most difficult to reduce among all sectors in this process. To reduce $\mathbf{3^+}\vec{e}$ in this sector, which contains 56 integrals, we need to set up 55 relations with $d_{\text{max}}=1$ and 1 relation with $d_{\text{max}}=5$.

%It took us several hours on a laptop to obtain these completely analytic relations.
%The relations have been checked by \verb"FIRE5"~\cite{Smirnov:2014hma} on several phase space points numerically. And the results are totally agreed.

For the second example, we show all nonplanar five-leg topologies in Fig.\ref{fig:5g}, and summarize corresponding reduction information in Table \ref{tab:time}. Among them, the most complicated one is the seven-propagator sector represented by Fig.\ref{fig:5g}(c). To reduce $\mathbf{3^+}\vec{e}$ in this sector, which contains 84 integrals, we need to set up 14 relations with $d_{\text{max}}=0$, 64 relations with $d_{\text{max}}=1$, 4 relations with $d_{\text{max}}=2$, and 2 relations with $d_{\text{max}}=4$.

\begin{table}[h]
	\centering
	\begin{tabular}{|c|c|c|c|c|c|c|c|}\hline
		\multicolumn{4}{|c|}{$g+g\rightarrow H+H$} & \multicolumn{4}{|c|}{$g+g\rightarrow g+g+g$}\\
		\cline{1-8}
		Sector&~Type~&$d_{\text{max}}$&~$\mathbf{m^+}$~&Sector&~Type~&$d_{\text{max}}$&$\mathbf{m^+}$\\\hline
		1(a) &7-NP  &1  &$\mathbf{3^+}$&2(a) &8-NP     &1 &$\mathbf{3^+}$\\
		1(b) &7-P   &1  &$\mathbf{3^+}$&2(b) &8-NP     &3 &$\mathbf{3^+}$\\
		1(c) &6-NP  &5  &$\mathbf{3^+}$&2(c) &7-NP     &4 &$\mathbf{3^+}$\\
		1(d) &6-P   &4  &$\mathbf{2^+}$&2(d) &6-NP     &2 &$\mathbf{3^+}$\\\hline
	\end{tabular}
	\caption{Main reduction information for sectors shown in Figs.\ref{fig:ggToHH} and \ref{fig:5g}. See text for details.}
	\label{tab:time}
\end{table}

Reduction relations for $\mathbf{m^+}\vec{e}$ of all sectors listed in Table \ref{tab:time} were obtained on a laptop with 4 core Intel i7-6500U CPU and 16GB of RAM within 1 day, and final analytical relations are available for download in electronic form from an ancillary file in the arXiv version. For a given phase space point, a given spacetime dimension $D$, and assuming that values of all simpler integrals are already known, then solving all these $\mathbf{m^+}\vec{e}$ by Gaussian elimination of obtained relations can be finished within 0.01 second, which should be efficient enough to do phase space integration. To compare with, \verb"FIRE5"~\cite{Smirnov:2014hma} needs several hours to reduce $\mathbf{m^+}\vec{e}$ to MIs at a given phase space point.
Though the final results obtained by \verb"FIRE5" are analytic in spacetime dimension $D$, we expect that it will not reach as high an efficiency as ours even if it works with a specific value of $D$.
We have checked point by point in phase space that relations obtained by our method agree with that obtained by \verb"FIRE5".  Technical details will be given in a forthcoming paper~\cite{liuma}.

%We can draw two main conclusions from the data: 1). The reductions for non-planar diagrams are usually more difficult than planar diagrams. 2). For diagrams with the same type, the more propagators, the easier the reduction is. The first nature is in accordance with our expectation, because non-planar sectors usually have more master integrals than planar sectors for their more complex structure. While the second nature can be understood by the completeness of the propagators. Higher completeness usually means easier reduction.

%%%%%%%%%%%%%%%%%%%%%%%%%%%%%%%%%%%%%%%%%%%%%%%%%
\sect{Summary and outlook}
In this article, we propose a new representation for Feynman integrals, which is defined by analytical continuation of an asymptotic series containing only vacuum integrals. The new representation translates the problem of computing Feynman integrals to the problem of performing analytical continuations. This new perspective of Feynman integrals may be helpful to deepen our understanding of Feynman integrals and scattering amplitudes.

As an application of the new representation, we construct a systematic and efficient reduction method for multiloop Feynman integrals. Different from the traditional IBP reduction method based Laporta's algorithm, we never involve large coupled linear systems because our method reduces integrals step by step, similar to the one-loop case. Therefore, once reduction relations in our method are obtained, the numerical evaluation can be much more efficient than IBP reduction, especially when dealing with multiscale problems. With two two-loop cutting-edge examples, we find that our method is indeed very powerful to reduce multiloop multiscale Feynman integrals.

In our reduction method, the appearance of additional masses does not introduce too many difficulties, because we have already introduced effective masses for each propagator. Therefore, for instance, the reduction of two-loop integrals in the five-gluon scattering process with a massive quark loop and $gg\to t\bar{t}g$ process, which are exceedingly difficult problems in the view of IBP reduction, should be achievable based on our method.

%From the two examples, we note an interesting fact. To determine a layer of scalar integrals, we need only to involve very few but rather expensive high mass dimension relations. So, if one could find a more efficient way to set up these relations, the application of this method would be more perfect. In future, we will consider this problem more in details.

%%%%%%%%%%%%%%%%%%%%%%%%%%%%%%%%%%%%%%%%%%%%%%%%%
\sect{Acknowledgments}
We thank Kuang-Ta Chao, Feng Feng, Yu Jia, David Kosower, Roman Lee, Zhao Li, Xiaohui Liu, Ce Meng, Jian-Wei Qiu, Chen-Yu Wang and Yang Zhang for helpful discussions. This work is in part supported by the Recruitment Program of Global Youth Experts of China.

\emph{Note added}: Recently, several preprints appeared, e.g., \cite{Mistlberger:2018etf,Kosower:2018obg,Borowka:2018dsa,Boehm:2018fpv,Chawdhry:2018awn,DiVita:2018nnh,Abreu:2018rcw,Banerjee:2018lfq,Chicherin:2018mue,Mastrolia:2018uzb,Xu:2018eos,Davies:2018qvx,Badger:2018enw,Mandal:2018cdj,Mishima:2018olh,Kardos:2018uzy,Frellesvig:2019kgj}, which are aimed at solving two-loop cutting-edge problems with IBP reduction method but equipped with many advanced techniques. Even with these improvements of IBP reductions, our reduction method is still very competitive.

%%%%%%%%%%%%%%%%%%%%%%%%%%%%%%%%%%%%%%%%%%%%%%%%%
%%%%%%%%%%%%%%%%%%%%%%%%%%%%%%%%%%%%%%%%%%%%%%%%%
% references
\providecommand{\href}[2]{#2}\begingroup\raggedright\endgroup

%\bibliographystyle{utphysMa}
%\bibliography{bibTex}

\end{document}